\documentclass[prl,twocolumn,superscriptaddress,floatfix,10pt]{revtex4-1}
\usepackage{graphicx,bm}
\usepackage{amsmath}
\usepackage{amsfonts}
\usepackage{amssymb}
\usepackage{braket}
\usepackage{bm}
\usepackage{color}
\usepackage{xcolor}
\usepackage[%
colorlinks=true,
urlcolor=blue,
linkcolor=blue,
citecolor=blue
]{hyperref}

\begin{document}

\title{Exciton band topology in  spontaneous quantum anomalous Hall insulators: applications to twisted bilayer graphene}
\author{Yves H. Kwan}
\affiliation{Rudolf Peierls Centre for Theoretical Physics,  Clarendon Laboratory, Oxford OX1 3PU, UK}
\author{Yichen Hu}
\affiliation{Rudolf Peierls Centre for Theoretical Physics,  Clarendon Laboratory, Oxford OX1 3PU, UK}
\author{Steven H. Simon}
\affiliation{Rudolf Peierls Centre for Theoretical Physics,  Clarendon Laboratory, Oxford OX1 3PU, UK}
\author{S. A. Parameswaran}
\affiliation{Rudolf Peierls Centre for Theoretical Physics,  Clarendon Laboratory, Oxford OX1 3PU, UK}

\begin{abstract}
We uncover topological features of neutral particle-hole pair excitations of correlated quantum anomalous Hall (QAH)  insulators whose approximately flat conduction and valence bands have equal and opposite non-zero Chern number.  Using an exactly solvable  model we show that the underlying band topology  affects both the center-of-mass and relative motion of particle-hole bound states. This leads to the formation of topological exciton bands whose features are robust to nonuniformity of both the dispersion and the Berry curvature. We apply these ideas to recently-reported  broken-symmetry spontaneous QAH insulators  in substrate aligned magic-angle twisted bilayer graphene. 
\end{abstract}
\date{\today}

\maketitle

\textit{Introduction.---}The structure of the ground states of condensed matter systems intricately affects their low-temperature properties, and is imprinted in the spectrum of low-energy quasiparticles and long-wavelength collective excitations.
Electrical insulators generically break no continuous symmetries and hence lack gapless collective modes, and are gapped to  charge transport in their bulk. However, interactions can stabilize neutral excitons, bound states of  a hole in the valence band and an electron in the conduction band. While also gapped, excitons  typically have lower energy than charged  excitations and  dominate the optical response of direct-gap semiconductors, where they can be excited at zero wavevector. 
More generally, excitons form at a fixed `center of mass' (CM)  wavevector $\bm{q}$ set by the momentum separation between the valence band maximum and conduction band minimum. The  spectrum and transport properties of excitons can also be modified by the topology of the electronic bands near these extrema~\cite{srivastava2015,zhou2015}. This is captured by the \emph{excitonic} Berry curvature~\cite{yao2008} linked to the evolution of the two-particle bound state across its Brillouin zone (BZ). Such considerations  are  relevant, for example, to two-dimensional transition-metal dichalcogenides~\cite{wang2018}, where the valley-contrasting anomalous velocity of excitons has been experimentally observed~\cite{onga2017}.

Here, we focus on excitons in correlated insulators formed in moir\'e heterostructures of twisted bilayer graphene (TBG) aligned with hexagonal boron nitride (hBN). In the `magic angle' regime, absent interactions, the relevant band structure  has a gapped Dirac dispersion with four degenerate bands
below and above charge neutrality~\cite{bultinck2019,zhang2019b}.
Members of each degenerate quartet are labeled by   spin ($\sigma =\uparrow, \downarrow$) and valley ($\tau = \pm$) indices. The valleys correspond to the $\pm K$ points of the single-layer  BZ, have  Chern numbers $C =\tau$, and are interchanged by time-reversal symmetry (TRS). At integer filling, the suppressed bandwidth ($\lesssim 10$\,meV) allows interactions to stabilize TRS-breaking valley- and spin-polarized states in which a partial subset of the bands is fully occupied  --- a mechanism proposed to explain the observed quantized anomalous Hall (QAH) response  in hBN-TBG~\cite{serlin2019}.

We identify several striking features of the exciton spectrum in hBN-TBG linked to the flatness and nontrivial Chern number of the underlying single-particle bands coupled with the spontaneous breaking of time-reversal and spin rotation symmetries. We root our  understanding of universal topological features in an analytically  tractable  model that mimics the features of the  hBN-TBG band structure by  leveraging the mapping between $|C|=1$ Chern bands and Landau levels (LLs). Our four-band model  has perfectly flat dispersion and uniform Berry curvature, and consists of spinful electron LLs whose Chern number has  a sign set by a two-fold degenerate valley index. 
We  examine  excitations of a fully spin-and-valley-polarized state with one filled LL.  
We show that the intravalley  spin-wave mode has the gapless quadratic dispersion expected for Goldstone modes  of a conserved order parameter, consistent with closely related quantum Hall ferromagnets~\cite{bychkov1981,kallin1984}. In striking  contrast, we  show that  the intervalley excitonic bands of our model are gapped and exactly flat. 
The flatness of the bands admits low-energy $\bm{q}=0$ excitons  throughout the BZ,
and also leads us to consider the dynamics of the excitonic CM which is conjugate to $\bm{q}$. Strikingly, we find that the CM motion experiences significant anomalous velocity, linked to the Berry curvature of the evolution  of the  particle-hole (PH) pair wavefunction as  $\bm{q}$ evolves across the CM.
We  demonstrate that these qualitative features survive the introduction of finite bandwidth and Berry curvature inhomogeneity, and discuss the results in a microscopic model of hBN-TBG. Our work  illustrates that correlated ground states in moir\'e heterostructures can host unconventional excitations, whose many-body physics  we explore elsewhere~\cite{kwan2020}.

\begin{figure}[t]
	\includegraphics[width=\linewidth,trim={0.0cm 19.2cm 6cm 0cm},clip=true]{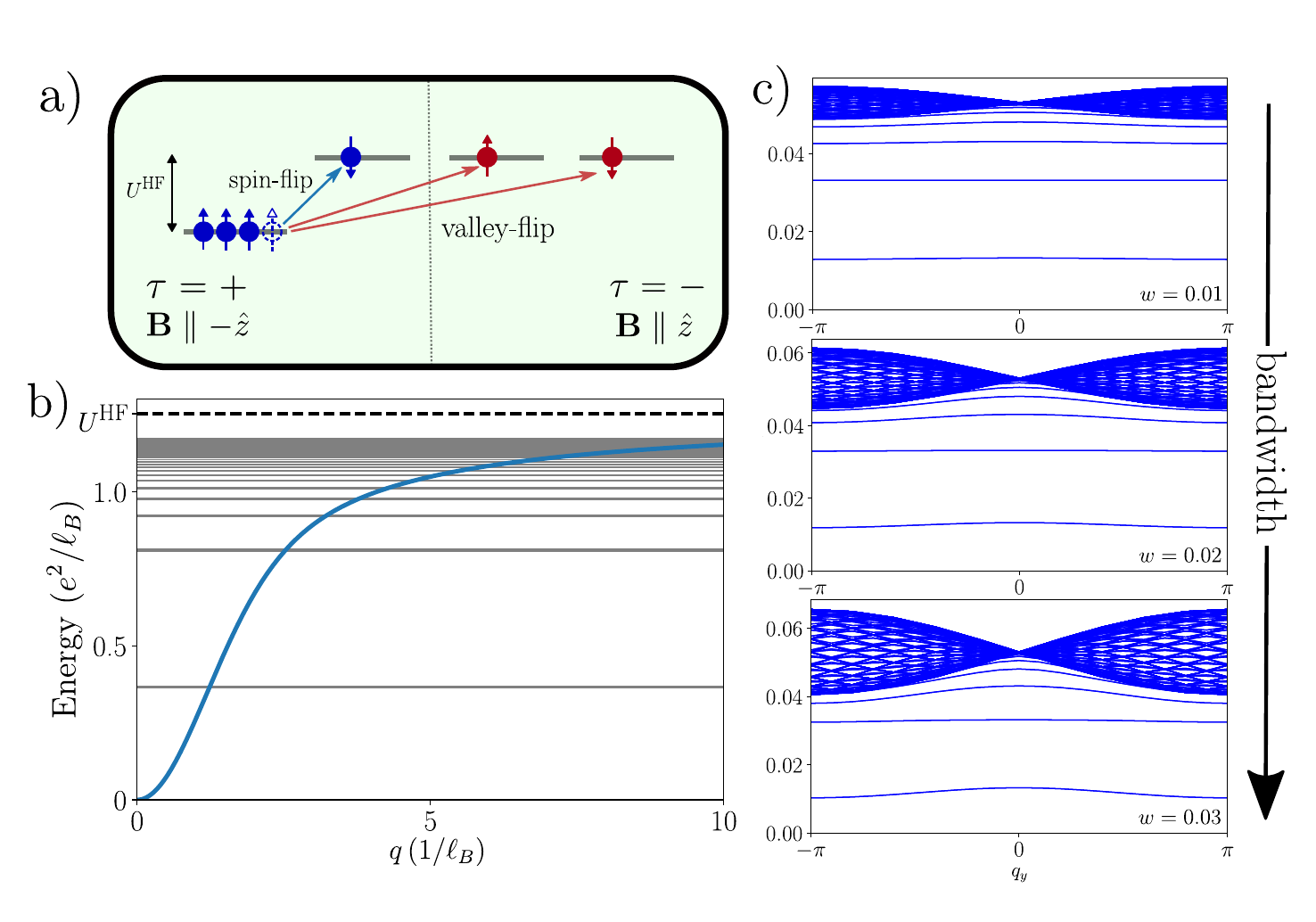}
	\caption{(a) Schematic of the four-band LLL model at $\nu=1$ showing the exchange-splitting $U^\text{HF}=\sqrt{\frac{\pi}{2}}\frac{e^2}{\ell_B}$ and the different neutral excitation types. (b) Spin wave energy  (blue curve) as a function of $y$-momentum $q$ for $\alpha=0$. Horizontal lines show the momentum-independent valley-flip exciton energies. Both modes saturate to $U^\text{HF}$. (c) Intervalley exciton spectrum at $q_x=0$ in the magnetic Brillouin with increasing bandwidth $w=0.01,0.02,0.03$. Calculations were performed on a $20\times20$ momentum-space mesh.}
	\label{FigSpinFlip}
	\end{figure}

\textit{Exactly solvable model.---} We exploit the topological equivalence between $|C|=1$ Chern bands and LLs, and 
consider a system of four flavors of electronic LLs confined to the plane (Fig.~\ref{FigSpinFlip}a). The two valleys $\tau=\pm$ experience opposite magnetic fields $\bm{B}=-\tau B\hat{z}$, chosen to model the Chern band structure of TBG, 
and we neglect Zeeman splitting since there is no {\it real} external magnetic field.
 In Landau gauge $\bm{A}_\tau=-\tau Bx\hat{y}$, the lowest Landau level (LLL) wavefunctions are 
$
\phi_{k\tau}(\bm{r})\propto e^{iky}e^{-\frac{(x-\tau k)^2}{2}}
$, which are created by $c^\dagger_{k\tau\sigma}$. We take $\ell_B\equiv (\hbar/eB)^{1/2} =1$ throughout.  The projected  LLL Hamiltonian is~\cite{karlhede1999}  
	\begin{align}\label{EqnLLLHamiltonian}
	\hat{H}&=\frac{1}{2}\sum_{\substack{kpq\\\tau\tau'\sigma\sigma'}}V_{\tau\tau'}(k,p,q)c^\dagger_{k+q,\tau\sigma}c^\dagger_{k+p-q,\tau'\sigma'}c_{k+p,\tau'\sigma'}c_{k,\tau\sigma}\nonumber\\
	&\,\,\,\,\,\,\,\,-\sum_{kp\tau\sigma}V_{\tau+}(k,p,0)c^\dagger_{k\tau\sigma}c_{k\tau\sigma}+\text{const}., 
	\end{align}
where the second line is a uniform background charge (equivalent to a filled  $\tau=+$ LL), and  $U(\bm{r})=e^2/r$ describes Coulomb interactions with LLL matrix elements   $
V_{\tau\tau'}(k,p,q)\equiv\bra{k+q,\tau;k+p-q,\tau'}\hat{U}\ket{k,\tau;k+p,\tau'}$. The form of~(\ref{EqnLLLHamiltonian})
is motivated by TBG, where interactions have  $SU(2)$ spin rotation invariance  and the suppressed intervalley scattering contributions are neglected~\cite{bultinck2019}.

We consider a uniform fully spin- and valley-polarized ground state, assuming without loss of generality that $(\tau, \sigma)= (+, \uparrow)$, viz. 
$\ket{G}\equiv\prod_{k}c^\dagger_{k+\uparrow}\ket{\text{vac}}$. The scenario with three filled flavors is equivalent via PH conjugation. Following Ref.~\cite{karlhede1999}, we compute the collective mode spectrum in the time-dependent Hartree-Fock approximation (TDHFA, equivalent to the generalized random phase approximation~\cite{nozieres1999}). To do so, we solve the dynamics restricted to the basis of single PH pairs, created by the neutral operators $b^\dagger_{\tau\sigma}(k,q)\equiv c^\dagger_{k+q,\tau\sigma}c_{k,+\uparrow}$ (where $q$ is the momentum transfer) that satisfy the equation of motion
\begin{align}\label{eq:TDHFA}
-i\partial_t&b^\dagger_{\tau\sigma}(k,q)=\big(\epsilon^\text{HF}_{\tau\sigma}(k+q)-\epsilon^\text{HF}_{+\uparrow}(k)\big)b_{\tau\sigma}^\dagger(k,q)\nonumber\\
&-\sum_{k'}V_{\tau+}(k+q,k'-k-q,k'-k)b_{\tau\sigma}^\dagger(k',q),
\end{align}
where $(\tau,\sigma)\neq(+,\uparrow)$, and the $k$-independent Hartree-Fock (HF) energies are $\epsilon^\text{HF}_{\tau\sigma}=-\delta_{\sigma\uparrow}\delta_{\tau+}\sum_{p}V_{++}(\cdot,p,p)$. Eq.~(\ref{eq:TDHFA}) is closed for a given $\tau$, $\sigma$, and $q$ as these are conserved by the Hamiltonian.
Thus,  TDHFA is {\it exact} for the one PH subspace when we neglect LL mixing. 

We solve (\ref{eq:TDHFA})  by  finding   operators $\gamma_{\tau\sigma}^\dagger(q)=\int dk\,\psi_{q\tau\sigma}(k)b^\dagger_{\tau\sigma}(k,q)$ such that to leading order in $\gamma_{\tau\sigma}(q)$ 
$[\hat{H}, \gamma^\dagger_{q\tau\sigma}(k)] =  \omega_{\tau\sigma}(q)  \gamma^\dagger_{q\tau\sigma}(k)$, where $\omega_{\tau\sigma}(q)$ is the excitation energy. The coefficients $\psi_{q\tau\sigma}(k)$ 
satisfy 
\begin{equation}\label{EqnFredholm}
\!\!\!\left(U^\text{HF}-\omega_{\tau\sigma}(q)\right)\psi_{q\tau\sigma}(k)=\int dk'\,T_{q,\tau}(k,k')\psi_{q\tau\sigma}(k'),\!\!\! 
\end{equation}
\normalsize
with kernel $T_{q,\tau}(k,k')=\frac{L_y}{2\pi}V_{\tau+}(k'+q,k-k'-q,k-k')$, where $U^\text{HF}=\sqrt{\frac{\pi}{2}}\frac{e^2}{\ell_B}$. Discretizing Eq. (\ref{EqnFredholm}) yields a 1D hopping problem for each $q$, with matrix element $T_{q,\tau}(k,k')$ between sites $k$, $k'$.

\textit{Intravalley spin-wave mode.---}
For $(\tau,\sigma)=(+,\downarrow)$ and fixed  $q$, the hopping kernel $T_{q,\tau}(k,k')$  depends only on $k-k'$. With the ansatz \mbox{$\psi_{q\tau\sigma}(k)\sim e^{ik\alpha}$}~\cite{karlhede1999}, the energy of this spin wave collective mode is given by~\cite{bychkov1981,kallin1984}
\begin{equation}
\omega_{+\downarrow}(q,\alpha)=U^{\text{HF}}-\int dk\,e^{ik\alpha}V_{++}(\cdot,k-q,k),
\end{equation}
plotted in Fig.~\ref{FigSpinFlip}a. The dispersion is isotropic in the $(q,\alpha)$ plane, and $\alpha$ can be interpreted as the $x$-momentum~\cite{karlhede1999}. $\omega_{+\downarrow}(q)$ is gapless and quadratic for $q\rightarrow 0$~\cite{bychkov1981,kallin1984,alavirad2019} and as $q\rightarrow \infty$ it saturates to $U^\text{HF}$ (the loss of exchange energy in creating a hole)  since in this limit electron and  hole are sufficiently distant that their Coulomb energy vanishes. 

\textit{Intervalley exciton mode.---}
A more interesting case is that of intervalley excitations where $\tau=-$. The spectrum is spin-independent since $\hat{H}$ is $SU(2)$-symmetric. In fact, {the spectrum $\omega_{-\sigma}(q)$ is also independent of $q$}, and hence macroscopically degenerate --- a consequence of the `shift symmetry' of the kernel, $T_{q,-}(k,k')=T_{q+2\delta,-}(k-\delta,k'-\delta)$~\cite{SupMat}, where increasing $q$ by $\delta$ corresponds to shifting the effective 1D hopping problem by $-\delta/2$. The sign and the factor of two is strongly suggestive of the notion that $b^\dagger_{\tau\sigma}(k,q)$ creates an excitation that couples to the magnetic field with an effective strength $2eB$ (recall the position-momentum locking of LLs, $\langle x\rangle=\tau kl_B^2$).  This leads to a discrete $q$-independent spectrum of excitonic bound states (Fig.~\ref{FigSpinFlip}a) described by harmonic oscillator wavefunctions
$\psi_{q\tau\sigma}(k;n)\propto H_n\big[\sqrt{2}(k+\frac{q}{2})\big]e^{-(k+\frac{q}{2})^2}$
centered at $-q/2$, where $H_n$ is a Hermite polynomial corresponding to LLs in an effective magnetic field $2B$.

For  rotationally invariant interactions we can capture key features of the excitons~\cite{SupMat,kwan2020}  by working in symmetric gauge and performing a PH transformation on the $\tau=+$ valley, yielding a two-body Hamiltonian for the $+$ hole and $-$ electron. The CM sector is a LLL problem for  a charge $-2e$ particle  in $-B\hat{z}$, yielding a  macroscopic degeneracy $N_{\text{exc}}=2N_{\Phi}$ of each valley-flip level due to the doubled coupling to the field. The relative motion corresponds to a charge $-e/2$ particle in the same field  and a Coulomb central potential. Solving this in terms of Haldane pseudopotentials~\cite{HaldaneSphere} yields discrete exciton binding energies
$E_m=-\frac{e^2}{\ell_B}\frac{\Gamma(m+\frac12)}{2\Gamma(m+1)}$ (where $m\geq0$ is an integer and $\Gamma$ is the gamma function),
in agreement with numerical solution of Eq.~\eqref{EqnFredholm}. The exciton is a composite neutral object which sees an effectively doubled magnetic field, and whose CM and relative motion are  topologically non-trivial (due  to Landau  quantization) for any interaction. 
 Semiclassical quantization also gives a macroscopic CM degeneracy and  discrete relative energy levels, because of the Lorentz-force deflection of electrons and holes as they attract in opposing magnetic fields. In contrast,  for the usual case of identical  fields the CM of the PH pair evolves in zero field and its energy is non-degenerate~\cite{GorkovDzyaloshkinskiiExciton}.

\textit{Perturbations away from the LL limit.---} We are  interested in studying Chern bands with small but non-zero dispersion and non-uniform Berry curvature. 
To model effects of the single-particle dispersion in our LL model, we transform to the magnetic Bloch basis indexed by two-dimensional momenta $\bm{k}$ in the magnetic BZ.  
Picking a square unit cell with side $a=\sqrt{2\pi}$ enclosing unit flux (for magic-angle TBG with $a\simeq 14$\,nm, this corresponds to $B\simeq 5$\,T), the single-particle magnetic Bloch operators are~\cite{bultinck2019} $d^\dagger_{\bm{k}\tau}=\frac{1}{\sqrt{N_x}}\sum_{n=0}^{N_x-1}e^{i\tau k_x(k_y+nQ)}c^\dagger_{k_y+nQ,\tau}$ where $Q=\frac{2\pi}{a}$ is the BZ length and the spin index has been dropped as we are focusing on intervalley modes. 
Following Ref.~\cite{bultinck2019} we introduce a potential $V(\bm{r})=-w(\cos(\frac{2\pi x}{a})+\cos(\frac{2\pi y}{a}))$, which is diagonal in this basis and projects to a single-particle dispersion \mbox{$\epsilon_{\bm{k}}=-we^{-\frac{\pi}{2}}(\cos k_xa +\cos k_ya)$} in the LL. 
Solving the discretized TDHFA equations, we find 
that exciton energies evolve  with the CM momenta $\bm{q}$, forming bands within the BZ (Fig.~\ref{FigSpinFlip}c). The topology of exciton bands is encoded in their Berry curvature~\cite{yao2008}, as we now summarize~\cite{SupMat}. The exciton state is~\footnote{The decomposition of the momenta into $\bm{k}\pm\bm{\frac{q}{2}}$ is required to properly decouple the relative and CM sectors.}
\begin{equation}\label{eq:excitonwf}
\ket{\psi^\text{exc}_{\bm{q}}}=\sum_{\bm{k}} \psi_{\bm{q}}(\bm{k})d^\dagger_{\bm{k+\frac{q}{2}},-}d^{\phantom\dagger}_{\bm{k-\frac{q}{2}},+}\ket{G}. 
\end{equation}
After PH-transforming the $+$ valley, we can write
\begin{equation}
\ket{u^\text{exc}_{\bm{q}}}=e^{-i\bm{q}\hat{\bm{R}}}\sum_{\bm{k}}\psi_{\bm{q}}(\bm{k})\ket{\phi^{\phantom*}_{\bm{k}+\bm{\frac{q}{2}},-}}\ket{\phi^*_{\bm{k}-\bm{\frac{q}{2}},+}},
\end{equation} where $\ket{\phi_{\bm{k},\tau}}$ are the single-particle Bloch states, and the  $e^{-i\bm{q}\hat{\bm{R}}}$ prefactor ensures that the cell-periodic part $\ket{u^\text{exc}_{\bm{q}}}$  of  $\ket{\psi^\text{exc}_{\bm{q}}}$ satisfies $\bm{q}$-independent boundary conditions~\cite{resta2000}.  The Berry connection and gauge-invariant Berry  curvature are then computed from $\ket{u^\text{exc}_{\bm{q}}}$.  If $\bm{a}^{\tau} = i\bra{u^\tau_{\bm{q}}} \bm{\nabla}_{\bm{q}} \ket{u^\tau_{\bm{q}}}$ and $f^{\tau}  = \bm{\nabla}_{\bm{q}}\times\bm{a}_{\tau}(\bm{q})$ are the Berry connection and curvature of the underlying single-particle bands, the exciton Berry curvature  takes the form
\begin{eqnarray}\label{EqnExcitonCurvature}
\Omega_{\text{exc}}(\bm{q}) =   \Omega_{\text{sp}}(\bm{q})+\Omega_{\text{e}}(\bm{q})+ \Omega_{\text{sp,e}}(\bm{q}),
\end{eqnarray}
where (defining $\bm{k}_\pm = \bm{k}\pm \frac{\bm{q}}{2}$) the first contribution
\begin{eqnarray}
\Omega_{\text{sp}}(\bm{q}) = \frac{i}{4} \sum_{\bm{k}} |\psi_{\bm{q}}(\bm{k})|^2\{f^+(\bm{k}_-)-f^-(\bm{k}_+)\}
\end{eqnarray}
 stems from the  single-particle Berry curvature,
\begin{eqnarray}
\Omega_{\text{e}}(\bm{q})\!&=&i\sum_{\bm{k}}\!\partial_{q_x}\!\psi_{\bm{q}}(\bm{k})\partial_{q_y}\!\psi^*_{\bm{q}}(\bm{k})\! - \!\partial_{q_y}\!\psi_{\bm{q}}(\bm{k})\partial_{q_x}\!\psi^*_{\bm{q}}(\bm{k})\,\,\,\,\,\,\,\,\end{eqnarray} 
 captures the BZ evolution of the
  envelope function, and
 \begin{eqnarray}
\Omega_{\text{sp,e}}(\bm{q})&=&\!\frac{i}{2}\!\sum_{\bm{k},\tau=\pm}\!\!\{ \partial_{q_y}|\psi_{\bm{q}}(\bm{k})|^2a^{-\tau}_{x}(\bm{k}_{\tau}) -  (x\leftrightarrow y)\}\,\,\,\,
 \end{eqnarray}
 describes the coupling between the envelope function and the  single-particle Berry connection.  (Due to the ambiguity in defining $\bm{a}$ and  the phase of $\psi_{\bm{k}}(\bm{q})$, only the combination $\Omega_{\text{e}}+\Omega_{\text{sp,e}}$ is gauge-invariant.)
   Numerically $\Omega_{\text{exc}}(\bm{q})$ is computed on a finite $k$-mesh by computing gauge-invariant (non-Abelian) lattice field strengths~\cite{fukui2005}. Integrating  $\Omega_{\text{exc}}(\bm{q})$ over the BZ gives a quantized exciton Chern number $C_\text{exc} =  \int_{\text{BZ}} \frac{d^2 q}{2\pi} \Omega_{\text{exc}}(\bm{q})$. 
	\begin{figure*}[t]
	\includegraphics[width=\linewidth,clip=true]{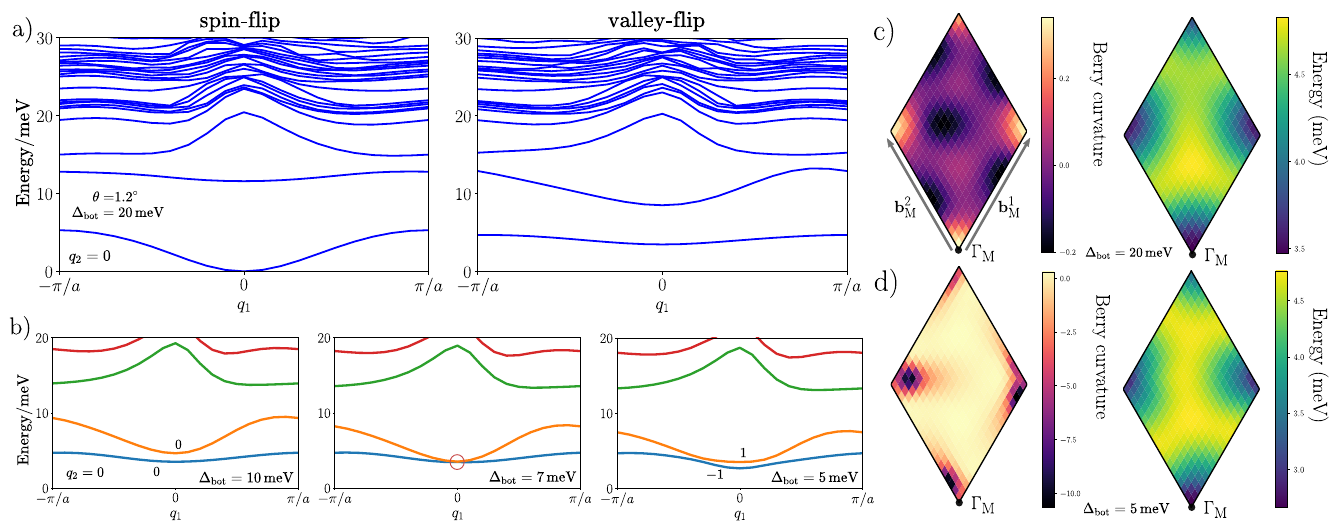}
	\caption{(a) Spin- and valley-flip exciton spectrum of $\nu=+3$ QAH state of TBG using the continuum model~\cite{bistritzer2011}, shown along the $K_{\text{M}}$-$\Gamma_{\text{M}}$-$K_{\text{M}}$ line in the moir\'e BZ (mBZ). (b) As substrate potential is varied, intervalley exciton bands cross at $\Gamma_{\text{M}}$ in a topological transition where  the Chern  numbers of the lowest bands change  as indicated with $|\Delta C_\text{exc}|=1$. 
	(c) Berry curvature (multiplied by the moir\'e BZ area) and exciton energy for the lowest valley-flip band of (a) in the moir\'e BZ. The origin $\Gamma_\text{M}$ and the reciprocal lattice vectors $\mathbf{b}^{1(2)}_\text{M}=\sqrt{3}k_\theta(\pm1/2,\sqrt{3}/2)$ are indicated, where the moir\'e wavevector $k_\theta=\frac{8\pi}{3\sqrt{3}a}\sin\frac{\theta}{2}$. (d) Same as (c) but for the $C_\text{exc}=-1$  band of the final panel of (b). System size is $20\times20$.}
	\label{FigTBG}
	\end{figure*}

 Armed with this definition  we return to our discussion of perturbing the solvable limit. At $w=0$ the bands are flat and two-fold degenerate~\footnote{For the simple cosine potential considered here, each pair of bands never fully detaches because $\epsilon_{\bm{k}}=\epsilon_{-\bm{k}}$.}, consistent with the CM experiencing a doubled effective field. As the bandwidth is increased, the upper levels merge into a continuum, which engulfs additional bands as $w$ grows (Fig.~\ref{FigSpinFlip}c). At large enough $w$ the lowest exciton band dips below $E=0$, signaling an instability to a partially-polarized state at the one-exciton level. We also introduce Berry curvature inhomogeneity by artificially deforming the Landau gauge states~\cite{SupMat}, and find that this leads to a weak exciton dispersion but preserves the Chern numbers. These results illustrate that the exciton dispersion  arises from  the interplay of the underlying band geometry, topology, dispersion, and interactions. We have also explicitly verified that low-lying exciton bands remain topological with $C_\text{exc}=1$ under these perturbations, even as they acquire dispersion and Berry curvature fluctuations of their own.
	
\textit{Microscopic calculation in TBG.---}
We now turn to  spin and valley-flip excitons of hBN-TBG, for which our starting point is the continuum model of Ref.~\cite{bistritzer2011} with twist $\theta\simeq1.2^\circ$ lying in the magic angle regime. We choose the inter-layer couplings  $w_{AA}=0.08$~eV and $w_{AB}=0.11$~eV to account for lattice relaxation effects~\cite{nam2017,carr2019}. The hBN alignment is introduced via a sublattice splitting $|\Delta|=20$~meV on the bottom layer. We use a dual-gate screened interaction $U(q)=\frac{e^2}{2\epsilon\epsilon_0 q}\tanh(qd_\text{sc})$ with relative permittivity $\epsilon=9.5$ and screening length $d_{\text{sc}}=40$~nm~\cite{bultinck2019b}, and account for interaction double-counting by measuring the density relative to that of decoupled graphene sheets at charge neutrality~\cite{xie2020,SupMat}.
Projecting to the eight central bands for simplicity, we calculate the fully flavor-polarized self-consistent QAH state at the experimentally-relevant filling $\nu=+3$~\cite{serlin2019}, from which we compute the single valley-flip or spin-flip excitonic spectra in Fig.~\ref{FigTBG}a~\cite{SupMat}. Consistent with previous studies that focused on energetics~\cite{wu2020,alavirad2019}, we find that the spin-flip mode is gapless and disperses quadratically at zero momentum, while the valley-flip mode is gapped. The  energetic separation and narrow bandwidth $\simeq1$~meV of the lowest valley-flip exciton band (Fig.~\ref{FigTBG}c) is promising for flat-band physics.

For the above parameters, 
we find that the lowest two exciton bands have $C_\text{exc}=0$. We emphasize however that the physics of TBG shows large sample-to-sample variations sensitive to the precise device parameters and experimental conditions. Indeed,  by varying the substrate strength, we can induce a set of band touching events which renders the lowest exciton band topological (Fig.~\ref{FigTBG}b,d). This reveals that the different terms in Eq.~\eqref{EqnExcitonCurvature} can give competing contributions to the exciton Berry curvature. Specifically, the non-trivial structure of the envelope function $\psi_{\bm{q}}(\bm{k})$ can render exciton bands trivial even if the underlying single-particle bands have equal and opposite Chern numbers and yield a nonzero gauge-invariant $\Omega_{\text{sp}}(\bm{q})$.  Despite these subtleties, it seems likely that hBN-TBG and other spontaneous QAH systems can host low-lying topological exciton branches in realistic parameter regimes. 

\textit{Discussion.---} We have studied the properties of excitons constituted of particles and holes from bands with equal and opposite Chern numbers, focusing on the Berry curvature experienced by the exciton center-of-mass momentum as  it evolves across the BZ. We first  studied  a solvable model and then showed that universal features are stable to including finite dispersion and Berry curvature inhomogenieties. Using these insights, we analysed the topology of the low-lying exciton dispersion in hBN-TBG. For realistic interactions we find  substantial exciton Berry curvature, integrating to a  non-zero Chern number for the lowest exciton band in a subset of the explored parameter space.

As with other topological collective modes~\cite{shindou2013,karzig2015,nalitov2015}, a non-zero Chern number for a bulk exciton band indicates the presence of chiral exciton modes~\cite{wu2017,gong2017,chen2017} localized at the boundary of the QAH domain, traversing the bulk gap to connect the band to one with a distinct Chern number. These modes allow unidirectional exciton transport, acting as chiral  channels for valley charge, but only emerge in TBG for a  narrow range of parameters.
However, even when the lowest exciton band has $C_{\text{exc}}=0$, we nevertheless find substantial curvature inherited from the underlying Chern bands (Fig.~\ref{FigTBG}c). 
This can drive anomalous exciton transport in the bulk~\cite{yao2008,onga2017}. Each valley-flip  exciton of QAH systems such as hBN-TBG is associated with a  $U(1)$  valley charge. Since the latter is to  very good approximation conserved in these systems,  excitons are likely long-lived. Direct optical addressing of  these excitons is challenged  by  the momentum mismatch between the valleys; however, it may be possible to supply this `missing momentum' from another source, e.g. phonons~\cite{Kukushkin1044}.
As conserved  bosons in a flat topological band, these valley-flip excitons are a potential platform for engineering neutral  bosonic  quantum Hall states, a question that we address in a companion work~\cite{kwan2020}.

\textit{Acknowledgments.---}
We thank N. Bultinck, G. Wagner, B. Lian, S.L. Sondhi, and M.P. Zaletel for useful discussions.
We acknowledge support from the European Research Council (ERC) under the European Union Horizon 2020 Research and Innovation Programme (Grant Agreement No.~804213-TMCS) and from EPSRC grant EP/S020527/1. Statement of compliance with EPSRC policy framework on research data: This publication is theoretical work that does not require supporting research data.

%

\onecolumngrid
\newpage

\section*{SUPPLEMENTARY INFORMATION FOR ``Exciton band topology in  spontaneous quantum anomalous Hall insulators: applications to twisted bilayer graphene''}
\begin{appendix}
	
	\section{Coulomb Matrix Elements}
	In this section we give expressions for the interaction matrix elements that enter the TDHFA equations for the collective modes of the lowest Landau levels in opposite fields. We define
	\begin{align}
	\begin{split}
	V_{\tau_1\tau_2\tau_3\tau_4}(k,p,q)&\equiv\bra{k+q,\tau_1;k+p-q,\tau_2}\hat{U}\ket{k,\tau_4;k+p,\tau_3}\\
	&=\iint d\bm{r}d\bm{r}'\,U(\bm{r}-\bm{r}')\phi^*_{k+q,\tau_1}(\bm{r})\phi^*_{k+p-q,\tau_2}(\bm{r}')\phi_{k+p,\tau_3}(\bm{r}')\phi_{k,\tau_4}(\bm{r}).
	\end{split}
	\end{align}
	where $U(\bm{r})=e^2/|\bm{r}|$ and $\phi_{k,\tau}(\bm{r})=\frac{1}{\sqrt{\pi^{\frac{1}{2}}L_y}}e^{iky}e^{-\frac{(x-\tau k)^2}{2}}$. After some work, this simplifies to
	\begin{gather}\label{EqnMatrixBesselFinalForm}
	V_{\tau_1\tau_2\tau_3\tau_4}(k,p,q)=\frac{e^2}{L_y}\sqrt{\frac{2}{\pi}}\int_{-\infty}^{\infty}dx\,K_0(|qx|)e^{-\frac{1}{2}(x^2+Bx+C)}\\
	B=k(-\tau_1+\tau_2+\tau_3-\tau_4)+p(\tau_2+\tau_3)+q(-\tau_1-\tau_2)\\
	\begin{align}
	C=&k^2\left[4-\frac{\left(\tau_1+\tau_2+\tau_3+\tau_4\right)^2}{4}\right]+p^2\left[2-\frac{\left(\tau_2+\tau_3\right)^2}{4}\right]+q^2\left[2-\frac{\left(\tau_1-\tau_2\right)^2}{4}\right]\\
	&+kp\left[4-\frac{\left(\tau_2+\tau_3\right)\left(\tau_1+\tau_2+\tau_3+\tau_4\right)}{2}\right]+kq\left[-\frac{\left(\tau_1+\tau_2+\tau_3+\tau_4\right)\left(\tau_1-\tau_2\right)}{2}\right]\\
	& +pq\left[-2-\frac{\left(\tau_2+\tau_3\right)\left(\tau_1-\tau_2\right)}{2}\right],
	\end{align}
	\end{gather}
	where $K_0$ is a modified Bessel function of the second kind. We now focus on the cases of relevance to density-density interactions considered in the main text.
	
	\subsection{$V_{++++}$}
	This is the only matrix element that contributes to the pure spin flip excitation. Here, $B=2(p-q)$ and $C=q^2+(p-q)^2$. Clearly $V_{++++}(k,p,q)$ is independent of $k$---this makes sense because $k$ is some overall momentum of the four states entering the matrix element, and this can always be removed by shifting the integrations of $x,x'$.
	\subsection{$V_{-++-}$}
	This appears in the direct Coulomb interaction between the constituent electron and hole of a intervalley exciton. Here, $B=4k+2p$ and $C=4k^2+p^2+q^2+4kp$. From these coefficients we can deduce the very important property
	\begin{equation}
	V_{-++-}(k,p,q)=V_{-++-}(k+\delta,p-2\delta,q).
	\end{equation}
	This is satisfied for any interaction, not just Coulomb, and leads to the macroscopic degeneracy of each exciton energy level as explained in the main text.

	\section{Symmetric Gauge Calculation}
	In this section we analyze the valley-flip collective mode problem in the symmetric gauge, which is simpler due to the rotational invariance of the interaction. (Of course the Landau gauge is more suited to investigating the effects of a periodic potential.) It also has the advantage of being more natural from the perspective of finite-size considerations (i.e. a Hall droplet). The single-particle states for the two valleys are
	\begin{gather}
	\chi_{m,+}(z)=\frac{A_m}{\sqrt{\pi}}z^me^{-\frac{|z|^2}{4}}\\
	\chi_{m,-}(z)=\frac{A_m}{\sqrt{\pi}}z^{*m}e^{-\frac{|z|^2}{4}}\\
	A_m=\left(2^mm!\right)^{-\frac{1}{2}},
	\end{gather}
	and the Hamiltonian is (spin has been removed) 
	\begin{gather}
	\hat{H}=\frac{1}{2}\sum_{m_1m_2n_1n_2\tau\tau'}V_{\tau\tau'}(n_1n_2m_2m_1)c^\dagger_{n_1\tau}c^\dagger_{n_2\tau'}c_{m_2\tau'}c_{m_1\tau}\\
	V_{\tau\tau'}(n_1n_2m_2m_1)\equiv\int d\bm{r}d\bm{r'}\,U_{\tau\tau'}(\bm{r}-\bm{r'})\chi_{n_1,\tau}^*(z)\chi_{n_2,\tau'}^*(z')\chi_{m_2,\tau'}(z')\chi_{m_1,\tau}(z).
	\end{gather}
	This is $(m_1,m_2)\rightarrow(n_1,n_2)$ scattering. Angular momentum conservation imposes selection rules that depend on the valleys involved. If $\tau=\tau'$, we have $n_1+n_2=m_1+m_2$. If $\tau\neq\tau'$ we have $n_1-n_2=m_1-m_2$. If the intervalley and intravalley interactions are the same, we have $V_{+-}(n_1n_2m_2m_1)=V_{++}(n_1m_2n_2m_1)$.
	
	Let $\ket{G}=\prod_mc^\dagger_{m+}\ket{\text{vac}}$ be the reference ground state. The basis of single PH states will be parameterized as 
	\begin{equation}
	\ket{a,b}\equiv c^\dagger_{a,-}c_{b,+}\ket{G}.
	\end{equation}
	
	Here we will fix intervalley and intravalley interactions to be the same Coulomb form -- assuming otherwise won't change the structure of the exciton states. From the selection rules, it is easy to see that $\ket{0,0}$ is an exact eigenstate with energy above $\ket{G}$ of
	\begin{equation}
	E_{0,0}=\sum_nV_{++}(0n0n)-V_{+-}(0000)=\frac{e^2}{l_B}\left(\sqrt{\frac{\pi}{2}}-\frac{\sqrt{\pi}}{2}\right),
	\end{equation}
	which matches the lowest exciton energy level described in the main text.
	
	In fact from angular momentum conservation, the seed basis state $\ket{n,0}$ only couples to $\ket{n,0},\ket{n-1,1},\ldots,\ket{0,n}$. The matrix elements are $\bra{n-l',l'}\hat{H}'\ket{n-l,l}=-V_{+-}(l,n-l',n-l,l')$, where the prime on $\hat{H}'$ indicates deduction of the ground state energy. 
	
	Solving the finite-dimensional problem generated by $\ket{n,0}$, we obtain the exciton binding energies in Eq.~\eqref{EqnCoulombExcitonEnergies} for $m=0,\ldots,n$. Furthermore upon examination of the eigenvectors of the lowest exciton state, we find that the coefficients are consistent with a first-quantized wavefunction $\sim z^{*n}e^{-\frac{|u|^2}{8}-\frac{|z|^2}{2}}$ (Eq.~\eqref{EqnSingleExcitonWF}), independent of the form of the interaction potential (as long as the $m=0$ pseudopotential is the largest one). This is consistent with the toy exciton calculation -- projection of the CM to the LLL allows for a polynomial prefactor $f(z^*)$ in the CM coordinate. 
	
\begin{figure}
		\includegraphics[trim={0cm 21.5cm 0cm 0cm}, width=1\linewidth,clip=true]{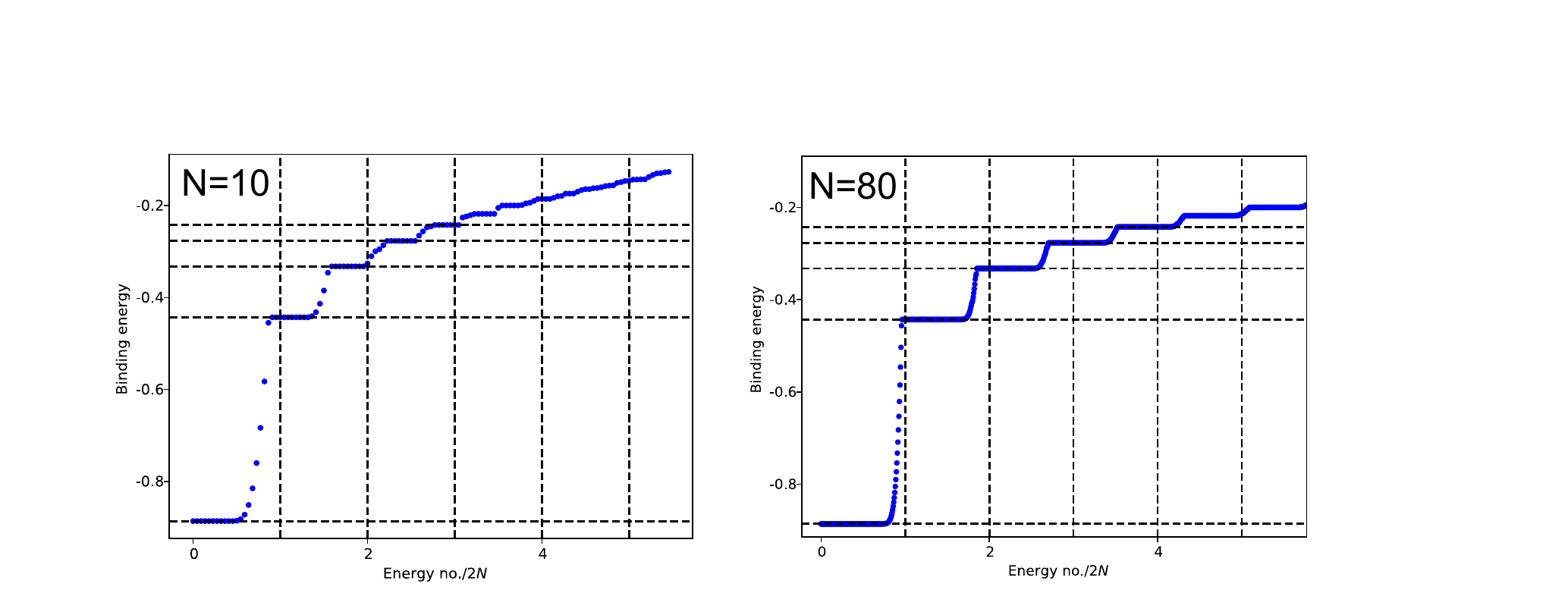}
		\caption{Ordered valley flip exciton energies in units of $\frac{e^2}{\ell_B}$ for a system that only allows PH pairs whose components have angular momentum $\leq N$. The vertical dashed lines are spaced every $2N$, which is the expected exciton degeneracy as $N\rightarrow\infty$. The horizontal dashed lines are the expected bulk binding energies (Eq.~\eqref{EqnCoulombExcitonEnergies}). We have zoomed in on the first $\sim5$ exciton energy levels.}
		\label{FigSymmetricGaugeFinite}
	\end{figure}

	The utility of doing the collective mode calculation in the symmetric gauge is that it provides a controlled way to count the degeneracy of exciton states as we approach the thermodynamic limit. For example, we can restrict ourselves to only allow particle-hole pairs whose components have angular momenta $\leq N$. Edge effects are apparent by considering say $\ket{N,N}$, for which there exist no other legal states to couple to. As $N\rightarrow\infty$ we expect to recover the physics of the bulk. Figure \ref{FigSymmetricGaugeFinite} shows that as $N$ increases, the exciton degeneracy approaches $2N$ per energy level---this makes intuitive sense since the exciton has twice the coupling to the magnetic field, and therefore has twice as many states (per relative motion configuration).
	
	\section{Electron-hole Two-body Problem}
	Consider an electron and a hole with charges $-e$ and $e$ respectively, and identical masses $m$. They are confined to the 2D $x-y$ plane, and are coupled to the vector potentials $\bm{A}_e=-\bm{A}_h=\bm{A}$, where the relative sign reflects the opposite magnetic fields felt by the two particles. They interact via an attractive potential $-v_d(r)$ with $v_d(r)>0$, so that the Hamiltonian is 
	\begin{equation}
	H_\text{exc}=\frac{(\bm{p}_e+e\bm{A})^2}{2m}+\frac{(\bm{p}_h+e\bm{A})^2}{2m}-v_d(r).
	\end{equation}
	
	Now transform to centre-of-mass (CM) and relative coordinates 
	\begin{gather}
	\bm{R}=\frac{\bm{r}_h+\bm{r}_e}{2},\quad \bm{r}=\bm{r}_h-\bm{r}_e\\
	\bm{P}=\bm{p}_h+\bm{p}_e,\quad \bm{p}=\frac{\bm{p}_h-\bm{p}_e}{2}.
	\end{gather}
	In this basis, the Hamiltonian cleanly decouples into CM and relative sectors
	\begin{gather}
	H_\text{exc}=H_{\bm{R}}+H_{\bm{r}}\\
	H_{\bm{R}}=\frac{(\bm{P}+2e\bm{A})^2}{4m}\\
	H_{\bm{r}}=\frac{(\bm{p}+\frac{e}{2}\bm{A})^2}{m}-v_d(r).
	\end{gather}
	
	It will be most convenient to use the symmetric gauge $\bm{A}=\frac{B}{2}(-y\hat{x}+x\hat{y})$. The CM is a Landau level problem for particles of mass $2m$ and charge $-2e$ in a magnetic field $B$. This should be projected to the LLL, which provides the macroscopic degeneracy of the exciton energy levels. Owing to the increased coupling $(-2eB)$ to the magnetic field, the degeneracy is $2N_\Phi$, twice that of the single-particle states.  
	
	On the other hand, the relative sector is mapped to a particle of mass $m/2$ and charge $-e/2$ in a magnetic field $B$, subject to an attractive central potential $-v_d(r)$. Upon projection to the LLL, $H_{\bm{r}}$ becomes easy to solve since the matrix elements of the potential are diagonal in the basis of LLL angular momentum eigenstates. Therefore the binding energies are simply Haldane pseudopotentials, and are obtained using symmetric gauge wavefunctions with magnetic length $\tilde\ell_B=\sqrt{2}\ell_B=\sqrt{2/eB}$
	\begin{equation}
	E_m=-\bra{\psi_{LLL}^m}v_d(r)\ket{\psi_{LLL}^m}=-\frac{1}{2^{m-1} m! \tilde\ell_B^{2(m+1)}}\int dr\,v_d(r)r^{2m+1}e^{-\frac{r^2}{2\tilde\ell_B^2}}.
	\end{equation}
	For the Coulomb interaction $v_d(r)=-e^2/r$, we obtain
	\begin{equation}\label{EqnCoulombExcitonEnergies}
	E_m=-\frac{e^2}{ \ell_B}\times\frac{\sqrt{\pi}}{2}\frac{(2m-1)!!}{2^m m!}.
	\end{equation}
	
	In terms of complex coordinates $z=\frac{z_h+z_e}{2}$ and $u=z_h-z_e$, the general form of an exciton eigenfunction in the $m$-th relative angular momentum channel is then 
	\begin{equation}\label{EqnSingleExcitonWF}
	\psi^m_{\text{exc}}\sim \tilde{f}(z^*)u^me^{-\frac{|z|^2}{2\ell_B^2}}e^{-\frac{|u|^2}{8\ell_B^2}}
	\end{equation}
	where $\tilde{f}(z)$ is anti-analytic in $z$.

	\section{Magnetic Bloch Basis}
	In this section we construct the magnetic Bloch basis -- this allows for an investigation into the effects of a periodic potential, as well as an explicit route to calculating the Berry curvature of the exciton bands.
	We take one flux per square unit cell  $a^2=2\pi \ell_B^2=2\pi$. Following Ref.~\cite{bultinck2019}, the relation between the Landau gauge basis $c_{k,\tau}$ and the magnetic Bloch basis $d_{\bm{k},\tau}$ on a system of size $L_x=N_xa,L_y=N_ya$ is
	\begin{gather}
	d^\dagger_{\bm{k},\tau}=\frac{1}{\sqrt{N_x}}\sum_{n=-\frac{N_x}{2}}^{\frac{N_x}{2}-1}e^{i\tau k_x(k_y+nQ)}c^\dagger_{k_y+nQ,\tau}\\
	c^\dagger_{k_y+nQ,\tau}=\frac{1}{\sqrt{N_x}}\sum_{k_x} e^{-i\tau k_x(k_y+nQ)}d^\dagger_{\bm{k},\tau}\\
	k_x=\frac{2\pi n_{k_x}}{N_xa},\quad n_{k_x}=0,\ldots,N_x-1\\
	k_y=\frac{2\pi n_{k_y}}{N_ya},\quad n_{k_y}=0,\ldots,N_y-1.
	\end{gather}
	where $Q=\frac{2\pi}{a}=a$ is the BZ side length. 
	
	The uniform Berry curvature of the SP states can be explicitly verified. The magnetic Bloch functions are
	\begin{equation}\label{EqnMagneticBlochFunction}
	\psi_{\bm{k},\tau}(\bm{r})=\frac{1}{\sqrt{N_x}}\sum_{n=-\frac{N_x}{2}}^{\frac{N_x}{2}-1}e^{i\tau k_x(k_y+nQ)}\phi_{k_y+nQ,\tau}(\bm{r}),
	\end{equation}
	which obeys the boundary conditions appropriate for magnetic translations: $\psi_{\bm{k},\tau}(\bm{r}+a\hat{y})=e^{ik_ya}\psi_{\bm{k},\tau}(\bm{r})$ and \mbox{$\psi_{\bm{k},\tau}(\bm{r}+a\hat{x})=e^{ik_xa}e^{i\tau Qy}\psi_{\bm{k},\tau}(\bm{r})$}. The cell-periodic part is
	\begin{equation}
	u_{\bm{k},\tau}(\bm{r})=e^{-i\bm{kr}}\psi_{\bm{k},\tau}(\bm{r})=\frac{1}{\sqrt{N_xL_y\pi^\frac{1}{2}}}\sum_ne^{i\tau k_x(k_y+nQ)}e^{-ik_xx}e^{inQy}e^{-\frac{[x-\tau (k_y+nQ)]^2}{2}},
	\end{equation}
	from which we may calculate the Berry connection and curvature
	\begin{gather}
	\bm{a}(\bm{k})=-i\braket{u(\bm{k})|\partial_{\bm{k}}|u(\bm{k})}=(0,\tau k_x)\\
	f=\partial_{k_x}a_y-\partial_{k_y}a_x=\tau.
	\end{gather}
	Therefore the band of LLL magnetic Bloch states (as well as higher LLs) has Chern number $C=\tau$. 
	
	\subsubsection{Interaction Hamiltonian}
	The position operator projected to the LLL is
	\begin{gather}
	\psi^\dagger_\tau(\bm{r})=\frac{1}{\sqrt{L_y\pi^{\frac{1}{2}}}}\sum_ke^{-iky}e^{-\frac{1}{2}(x-\tau k)^2}c^\dagger_{k\tau}\\
	k=\frac{2\pi n_k}{N_ya},\quad k=-\frac{N_x}{2},\ldots,\frac{N_x}{2}-1.
	\end{gather}
	The number density operator in momentum space projected to the LLL is
	\begin{align}
	n_\tau(\bm{q})&\equiv\int d\bm{r}\,e^{-i\bm{q}\bm{r}}\psi_\tau^\dagger(\bm{r})\psi_\tau(\bm{r})\\
	&=\int d\bm{r}\,e^{-i\bm{q}\bm{r}}\frac{1}{L_y\sqrt{\pi}}\sum_{k,k'}e^{i(k'-k)y-\frac{1}{2}(x-\tau k)^2-\frac{1}{2}(x-\tau k')^2}c^\dagger_{k\tau}c_{k'\tau}\\
	&=\int dx\,e^{-iq_xx}\frac{1}{\sqrt{\pi}}\sum_ke^{-\frac{1}{2}(x-\tau k)^2-\frac{1}{2}(x-\tau(k+q_y))^2}c^\dagger_{k\tau}c_{k+q_y,\tau}\\
	&=e^{-\frac{\bm{q}^2}{4}}\sum_ke^{-iq_x\tau(k+\frac{q_y}{2})}c_{k\tau}^\dagger c_{k+q_y,\tau}.
	\end{align}
	We now rewrite this in terms of the magnetic Bloch operators
	\begin{equation}
	n_\tau(\bm{q})=e^{-\frac{\bm{q}^2}{4}}\frac{1}{N_x}\sum_{\lfloor{k}\rfloor,n_k,k_x,k_x'}e^{-iq_x\tau(\lfloor{k}\rfloor+n_kQ+\frac{q_y}{2})}e^{-i\tau k_x(\lfloor{k}\rfloor+n_kQ)}e^{i\tau k_x'(\lfloor{k}\rfloor+n_kQ+q_y)}d^\dagger_{(k_x,\lfloor{k}\rfloor)\tau}d_{(k_x',\lfloor{k+q_y}\rfloor)\tau}
	\end{equation}
	Where $\lfloor k\rfloor$ means $k\,\,\text{mod}\,\,Q$ which resides inside the 1BZ (which is defined here as the square with corners at $(0,0)$ and $(Q,Q)$). We have used the decomposition $k=\lfloor{k}\rfloor+n_kQ$ where $n_k\in\mathbb{Z}$. Summing over $n_k$ leads to $N_x\delta(k_x'=\lfloor{q_x+k_x}\rfloor)$ since $k_x'$ is in the 1BZ by definition. Then
	\begin{align}
	n_\tau(\bm{q})&=e^{-\frac{\bm{q}^2}{4}}\sum_{\lfloor{k}\rfloor,k_x}e^{-iq_x\tau(\lfloor{k}\rfloor+\frac{q_y}{2})}e^{-i\tau k_x\lfloor{k}\rfloor}e^{i\tau\lfloor{q_x+k_x}\rfloor(\lfloor{k}\rfloor+q_y)}d^\dagger_{(k_x,\lfloor{k}\rfloor)\tau}d_{(\lfloor{k_x+q_x}\rfloor,\lfloor{k+q_y}\rfloor)\tau}\\
	&=e^{-\frac{\bm{q}^2}{4}}\sum_{\bm{k}\in\text{1BZ}}e^{i\tau\big(-q_x(k_y+\frac{q_y}{2})-k_xk_y+\lfloor{k_x+q_x}\rfloor(k_y+q_y)\big)}d^\dagger_{\bm{k}\tau}d_{\lfloor{\bm{k}+\bm{q}}\rfloor\tau}.
	\end{align}
	The interaction Hamiltonian is 
	\begin{align}
	\hat{H}_\text{int}&=\frac{1}{2}\sum_{\tau\tau'}\int d\bm{r}d\bm{r}'\,U_{\tau\tau'}(\bm{r}-\bm{r}')\psi_\tau^\dagger(\bm{r})\psi_{\tau'}^\dagger(\bm{r}')\psi_{\tau'}(\bm{r}')\psi_\tau(\bm{r})\\
	&=\frac{1}{2A}\sum_{\bm{q}\in\text{all}}\tilde{U}_{\tau\tau'}(\bm{q}):n_\tau(-\bm{q})n_{\tau'}(\bm{q}):
	\end{align}
	with $A=L_xL_y$ the area. Summation over $\tau,\tau'$ is implied. We substitute the density operators below. In the second equality, we split $\bm{q}\in\text{all}$ into a piece $\bm{q}\in\text{1BZ}$ and a reciprocal lattice vector $\bm{G}$, and define $\bm{Q}\equiv\bm{q}+\bm{G}$ (this should not be confused with $Q=\frac{2\pi}{a}$)
	\begin{align}
	\hat{H}_\text{int}=&\frac{1}{2A}\sum_{\bm{q}\in\text{all}}\sum_{\bm{k},\bm{k}'\in\text{1BZ}}\tilde{U}_{\tau\tau'}(\bm{q})e^{-\frac{\bm{q}^2}{2}}e^{i\tau\big(q_x(k_y-\frac{q_y}{2})-k_xk_y+\lfloor{-q_x+k_x}\rfloor(k_y-q_y)\big)}e^{i\tau'\big(-q_x(k_y'+\frac{q_y}{2})-k_x'k_y'+\lfloor{q_x+k_x'}\rfloor(k_y'+q_y)\big)}\\
	&\quad\quad\times d^\dagger_{\bm{k}\tau}d^\dagger_{\bm{k}'\tau'}d_{\lfloor{\bm{k}'+\bm{q}}\rfloor\tau'}d_{\lfloor{\bm{k}-\bm{q}}\rfloor\tau}\\
	=&\frac{1}{2A}\sum_{\bm{k}\bm{k}'\bm{q}\in\text{1BZ}}\sum_{\bm{G}}\tilde{U}_{\tau\tau'}(\bm{Q})e^{-\frac{\bm{Q}^2}{2}}e^{i\tau\big(Q_x(k_y-\frac{Q_y}{2})-k_xk_y+\lfloor{-q_x+k_x}\rfloor(k_y-Q_y)\big)}e^{i\tau'\big(-Q_x(k_y'+\frac{Q_y}{2})-k_x'k_y'+\lfloor{q_x+k_x'}\rfloor(k_y'+Q_y)\big)}\\
	&\quad\quad\times d^\dagger_{\bm{k}\tau}d^\dagger_{\bm{k}'\tau'}d_{\lfloor{\bm{k}'+\bm{q}}\rfloor\tau'}d_{\lfloor{\bm{k}-\bm{q}}\rfloor\tau}\\
	\equiv&\frac{1}{2}\sum_{\bm{k}\bm{k}'\bm{q}\in\text{1BZ}}F_{\tau\tau'}(\bm{k},\bm{k}',\bm{q})d^\dagger_{\bm{k}\tau}d^\dagger_{\bm{k}'\tau'}d_{\lfloor{\bm{k}'+\bm{q}}\rfloor\tau'}d_{\lfloor{\bm{k}-\bm{q}}\rfloor\tau}
	\end{align}
	Relabelling momenta, we obtain
	\begin{gather}
	\hat{H}_\text{int}=\frac{1}{2}\sum_{\bm{k}\bm{p}\bm{q}\in\text{1BZ}}M_{\tau\tau'}(\bm{k},\bm{p},\bm{q})d^\dagger_{\lfloor{\bm{k}+\bm{q}}\rfloor\tau}d^\dagger_{\lfloor{\bm{k}+\bm{p}-\bm{q}}\rfloor\tau'}d_{\lfloor{\bm{k}+\bm{p}}\rfloor\tau'}d_{\bm{k}\tau}\\
	M_{\tau\tau'}(\bm{k},\bm{p},\bm{q})=F_{\tau\tau'}(\lfloor{\bm{k}+\bm{q}}\rfloor,\lfloor{\bm{k}+\bm{p}-\bm{q}}\rfloor,\bm{q})\\
	F_{\tau\tau'}(\bm{k},\bm{k}',\bm{q})\equiv\frac{1}{A}\sum_{\bm{G}}\tilde{U}_{\tau\tau'}(\bm{Q})e^{-\frac{\bm{Q}^2}{2}}e^{i\tau\big(Q_x(k_y-\frac{Q_y}{2})-k_xk_y+\lfloor{-q_x+k_x}\rfloor(k_y-Q_y)\big)}e^{i\tau'\big(-Q_x(k_y'+\frac{Q_y}{2})-k_x'k_y'+\lfloor{q_x+k_x'}\rfloor(k_y'+Q_y)\big)}\label{EqnFmatrix}				
	\end{gather}
	For an initial state $\ket{G}$ fully polarized in $\tau=+$, the TDHFA equations for the collective mode \mbox{$\ket{\text{exc},\bm{q}}=\psi_{\bm{q}}(\bm{k})d^\dagger_{\lfloor{\bm{k}+\bm{q}}\rfloor,-}d_{\bm{k},+}\ket{G}$} are then
	\begin{align}\label{EqnTDHFALLLMagnetic}
	\omega_n(\bm{q})\psi_{\bm{q}n}(k)=&\sum_{\bm{p}}M_{++}(\bm{k},\bm{p},\bm{p})\psi_{{\bm{q}}n}(\bm{k})-\sum_{\bm{k}'}M_{\tau+}(\lfloor{\bm{k}'+\bm{q}}\rfloor,\lfloor{\bm{k}-\bm{k}'-\bm{q}}\rfloor,\lfloor{\bm{k}-\bm{k}'}\rfloor)\psi_{\bm{q}n}(\bm{k}')\\
	=&\sum_pF_{++}(\lfloor{\bm{k}+\bm{p}}\rfloor,\bm{k},\bm{p})\psi_{\bm{q}n}(\bm{k})-\sum_{\bm{k}'}F_{\tau+}(\lfloor{\bm{k}+\bm{q}}\rfloor,\bm{k}',\lfloor{\bm{k}-\bm{k}'}\rfloor)\psi_{\bm{q}n}(\bm{k}')
	\end{align}
	where $n$ is the band index. 
	
	\subsubsection{Berry Curvature of Valley-flip Exciton Bands}
	In the language of the magnetic Bloch basis, we can analytically compute the Berry curvature of the valley-flip exciton, since we know the analytical form of the envelope wavefunction. Recall the the form of the exciton creation operator $\gamma_{q,m}^\dagger$ for the $m$-th degenerate exciton level in the Landau gauge
	\begin{gather}
	\gamma^\dagger_{q,m}=\int dk\,\psi_{q,m}(k)b^\dagger(k,q)\\
	b^\dagger(k,q)=c^\dagger_{k+q,-}c_{k,+}\\
	\psi_{q,m}(k)\propto H_m\big[\sqrt{2}(k+\frac{q}{2})\big]e^{-(k+\frac{q}{2})^2}=\tilde{\psi}_m(k+\frac{q}{2})
	\end{gather}
	where $H_m$ is the $m$-th Hermite polynomial. $\tilde{\psi}(k)$ is centered at the origin with mean zero. Now consider following magnetic Bloch operator for the exciton 
	\begin{equation}
	\Gamma_{\bm{q},m}^\dagger\equiv\frac{1}{\sqrt{N_x}}\sum_n \gamma^\dagger_{q_y+2nQ,m}e^{-\frac{iq_x}{2}\left(q_y+2nQ\right)}
	\end{equation}
	whose wavefunction is (after performing a PH transformation in the $\tau=+$ band) \begin{equation}\psi^{\text{exc}}_{\bm{q},m}(\bm{r}_e,\bm{r}_h)=\frac{1}{\sqrt{N_x}}\sum_n\int dk\,\tilde{\psi}_{m}\left(k\right)e^{-\frac{iq_x}{2}(q_y+2nQ)}\phi_{k+\frac{q_y}{2}+nQ,-}(\bm{r}_e)\phi^*_{k-\frac{q_y}{2}-nQ,+}(\bm{r}_h).
	\end{equation}
	To verify that this is the right candidate, we check the magnetic Bloch theorem by shifting the exciton CM coordinate $\bm{R}=\frac{\bm{r}_e+\bm{r}_h}{2}$
	\begin{gather}
	\psi^{\text{exc}}_{\bm{q},m}(\bm{r}_e+a\hat{x},\bm{r}_h+a\hat{x})=e^{-i(2Q)\frac{y_e+y_h}{2}}e^{iq_x a}\psi^{\text{exc}}_{\bm{q},m}(\bm{r}_e,\bm{r}_h)\\
	\psi^{\text{exc}}_{\bm{q},m}(\bm{r}_e+\frac{a}{2}\hat{y},\bm{r}_h+\frac{a}{2}\hat{y})=e^{iq_y\frac{a}{2}}\psi^{\text{exc}}_{\bm{q},m}(\bm{r}_e,\bm{r}_h).
	\end{gather}
	In the first line, the factor of two in the phase $e^{-i(2Q)\frac{y_e+y_h}{2}}$ reflects the fact that the exciton has twice the coupling to the magnetic field (compare with the SP case earlier). Therefore, the magnetic unit cell shrinks by a factor of two in the $y$-direction, so that the magnetic BZ of the exciton doubles in area. 
	
	The `cell-periodic' part of the Bloch function is $u^\text{exc}_{\bm{q}}(\bm{r}_e,\bm{r}_h)=\exp(-i\bm{q}\cdot\frac{\bm{r}_e+\bm{r}_h}{2})\psi^\text{exc}_{\bm{q}}(\bm{r}_e,\bm{r}_h)$, and we can straightforwardly compute the Berry connections
	\begin{equation}
	-i\bra{u_{\bm{q}}}\partial_{q_x}\ket{u_{\bm{q}}}=0,\quad -i\bra{u_{\bm{q}}}\partial_{q_y}\ket{u_{\bm{q}}}=-\frac{q_x}{2}
	\end{equation}
	leading to a curvature $f_{xy}=-\frac{1}{2}$. Integrated over the doubled magnetic BZ, this leads to $C=-1$.
	\begin{figure}
		\includegraphics[trim={0cm 21.5cm 5.2cm 0cm}, width=0.8\linewidth,clip=true]{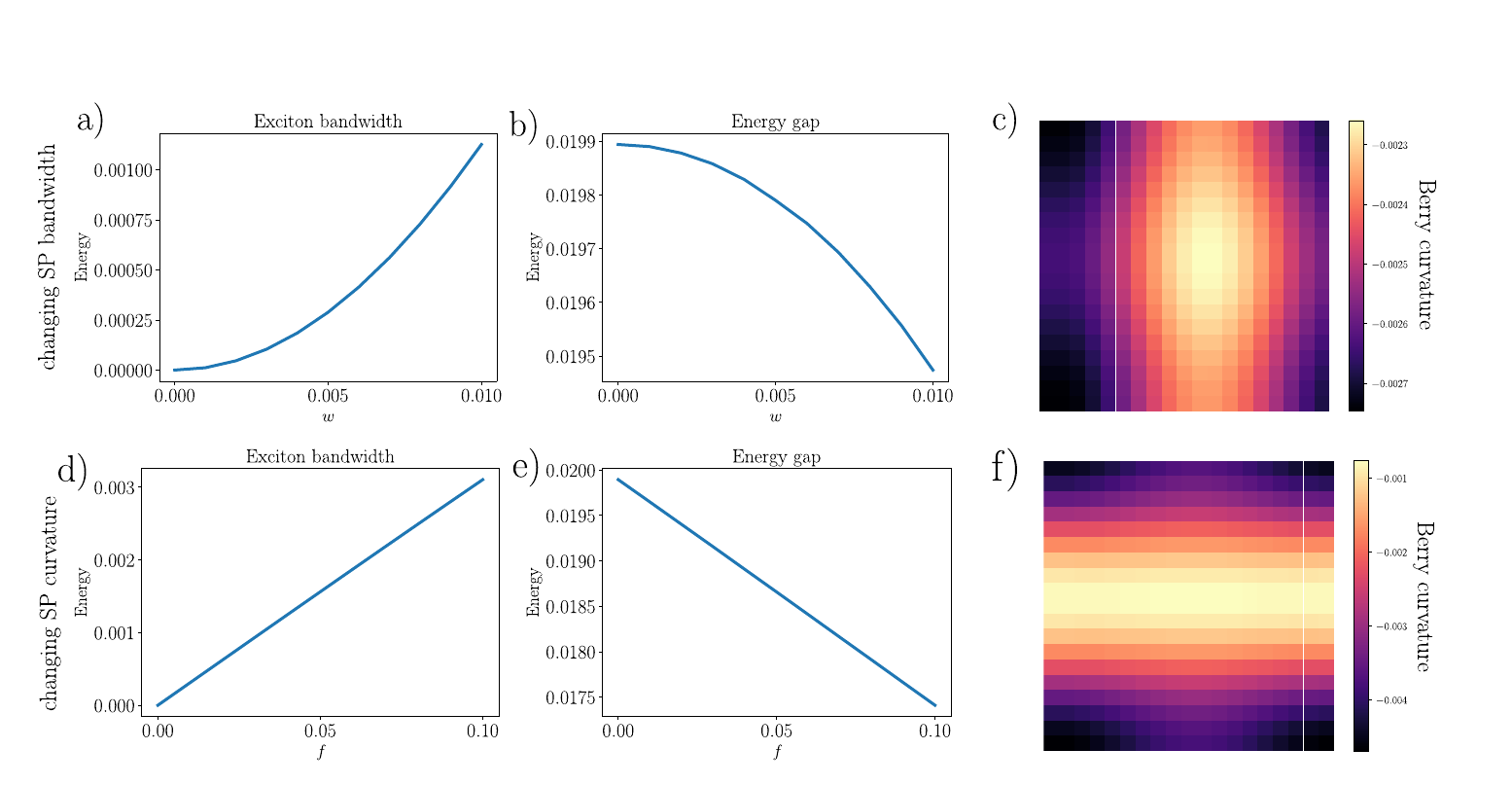}
		\caption{a) Bandwidth of lowest exciton band with as a function SP bandwidth parameterized by $w$. The SP potential is chosen as $V_p(x,y)= -w(\cos(\frac{2\pi x}{a})+\cos(\frac{2\pi y}{a}))$. b) Band gap between the first and second exciton bands with $w$. c) Berry curvature (normalized so that the mBZ sum over all cells is quantized in units of $2\pi$) at $w=0.005$. d-f) Same as a-c) except that the parameter being changed is the curvature of the underlying SP bands. The distortion of the bands used is $w(k_y)=f\sin(k_y/Q)$.}
		\label{FigPerturbations}
	\end{figure}	
	
	\subsubsection{Periodic Potential}
	Consider applying a potential $V_p(x,y)$ that is periodic in the square lattice. This enters the Hamiltonian as 
	\begin{equation}\label{EqnPerPotential}
	\hat{H}^\text{SP}_\tau=\int d\bm{r}\,V_p(x,y)\psi^\dagger_\tau(\bm{r})\psi_\tau(\bm{r}).
	\end{equation}
	Note that we restrict ourselves to identical potentials on both valleys, to preserve the time-reversal relation between the two. Since we are in the LLL, any periodic potential is diagonal in the magnetic Bloch basis. In particular, harmonics of the potential are mapped to the following dispersions
	\begin{align}
	\cos(\frac{2\pi n_xx}{a}+\phi_x)&\rightarrow \epsilon_{\bm{k},\tau}=e^{-\frac{n_x^2\pi}{2}}\cos(\tau n_xk_ya+\phi_x)\\
	\cos(\frac{2\pi n_yy}{a}+\phi_y)&\rightarrow \epsilon_{\bm{k},\tau}=e^{-\frac{n_y^2\pi}{2}}\cos(\tau n_yk_xa+\phi_y)
	\end{align}
	For the TDHFA, this perturbation will appear on the RHS of \ref{EqnTDHFALLLMagnetic} as a diagonal contribution of $\epsilon_{\lfloor{\bm{k}+\bm{q}}\rfloor,-}-\epsilon_{\bm{k},+}$. In Fig.~\ref{FigPerturbations}a-c, we have used the potential $V_p(x,y)= -w(\cos(\frac{2\pi x}{a})+\cos(\frac{2\pi y}{a}))$. The effect of the potential is to induce dispersion in the previously flat bands, as well as to redistribute the exciton Berry curvature. However the bands are still topological for finite $w$.
	
	\subsubsection{Distorted Landau Level States}
	To introduce Berry curvature inhomogeneity, we consider distorting the Landau gauge eigenstates with a `wobble' factor respecting the square lattice periodicity $w(k+Q)=w(k)$, and the time-reversal relation of the two valleys $w(-k)=-w(k)$
	\begin{equation}
	\phi_{k,\tau}(w;\bm{r})=\frac{1}{\sqrt{L_y\pi^{\frac{1}{2}}}}e^{iky}e^{-\frac{[x-\tau (k+w(k))]^2}{2}}.
	\end{equation}
	From this the distorted magnetic Bloch can be defined in analogy to Eq.~\eqref{EqnMagneticBlochFunction}. This perturbs the uniform Berry curvature in the $y$-direction, as can be explicitly verified
	\begin{gather}
	\bm{a}(\bm{k})=-i\braket{u(w;\bm{k})|\partial_{\bm{k}}|u(w;\bm{k})}=(-\tau w(k_y),\tau k_x)\\
	f=\partial_{k_x}a_y-\partial_{k_y}a_x=\tau+\tau \partial_{k_y}w(k_y),
	\end{gather}
	but leaves the Chern number unchanged at $C=\tau$.
	
	The TDHFA equations are the same except that the interactions are altered such that $F$ in Eq.~\eqref{EqnFmatrix} is replaced with
	\begin{align}
	\begin{split}
	F_{\tau\tau'}(\bm{k},\bm{k'},\bm{q})\equiv&\frac{1}{A}\sum_{\bm{G}}\tilde{U}_{\tau\tau'}(\bm{Q})e^{-\frac{Q_x^2}{2}}\exp\left[\frac{-1}{4}\left(\left[Q_y+w(k_y)-w(k_y-q_y)\right]^2+\left[Q_y+w(k_y'+q_y)-w(k_y')\right]^2\right)\right]\\
	\quad\times&\exp\left[i\tau\left(Q_x\left[\frac{2k_y-Q_y+w(k_y)+w(k_y-q_y)}{2}\right]-k_xk_y+\lfloor{k_x-q_x}\rfloor(k_y-Q_y)\right)\right]\\
	\quad\times&\exp\left[i\tau'\left(-Q_x\left[\frac{2k'_y+Q_y+w(k'_y)+w(k'_y+q_y)}{2}\right]-k'_xk'_y+\lfloor{k'_x+q_x}\rfloor(k'_y+Q_y)\right)\right].
	\end{split}
	\end{align}
	In Fig.~\ref{FigPerturbations}d-f, we have used $w(k_y)=f\sin(k_y/Q)$. The effect of the distortion is to induce dispersion in the previously flat bands, as well as to redistribute the exciton Berry curvature. However the bands are still topological for finite $f$.

	\section{Spin-Flip and Valley-Flip Excitons in TBG}
	In this section, we describe the equations for computing the spin-flip and valley-flip exciton bands in TBG. Our approach is equivalent to diagonalizing in the subspace of single PH pairs in the relevant symmetry sector. We do not consider `inter-band' excitons that are not charged under valley or spin. 
	
	We begin with a set of self-consistent Hartree-Fock (HF) band operators $d^\dagger_{\bm{k},\tau\sigma a}$ associated with a QAH state $\ket{\text{HF}}$ at $\nu=+3$, where $\tau,\sigma$ refer to valley and spin. Since our `active' subspace of the continuum model~\cite{bistritzer2011} comprises the eight central bands, the HF band index takes values $a=+,-$. Without loss of generality, we assume that $\ket{\text{HF}}$ is flavor polarized such that $d^\dagger_{\bm{k},K'\downarrow +}$ is the operator for the single unfilled band. We defer to the Supplementary Materials of Ref~\cite{kwan2020domain} for computational details of the continuum model, matrix elements, and HF procedure. 
	
	We first focus on intervalley excitons. Following the notation of Ref~\cite{ring2004}, we parameterize the exciton creation operator at momentum $\bm{q}$ as
	\begin{equation}
	Q^\dagger_\nu(\bm{q})=\sum_{\bm{k}a}X^\nu_{\bm{k}a}(\bm{q})d^\dagger_{\bm{k}+\bm{q},K'\downarrow+}d^{\phantom{\dagger}}_{\bm{k},K\downarrow a},
	\end{equation}
	where $\nu$ labels the exciton branch. The exciton envelopes $X^\nu_{\bm{k}a}(\bm{q})$ and energies $\omega_\nu(\bm{q})$ are obtained by solving the eigenvalue equation
	\begin{equation}
	\sum_{\bm{k'}a'}A_{\bm{k}a;\bm{k'}a'}(\bm{q})X^\nu_{\bm{k'}a}(\bm{q})=\omega_\nu(\bm{q})X^\nu_{\bm{k}a}(\bm{q}).
	\end{equation}
	
	The matrix $A_{\bm{k}x;\bm{k'}y}(\bm{q})$ can be split into a `single-particle' and an interaction piece. The single-particle contribution is
	\begin{gather}
	A^{\text{SP}}_{\bm{k}x;\bm{k'}y}(\bm{q})=\delta_{\bm{k}\bm{k'}}(H^\text{SP}_{\bm{k}+\bm{q},K'\downarrow; ++}\delta_{xy}-H^\text{SP}_{\bm{k},K\downarrow;yx})
	\end{gather}
	where the effective single-particle Hamiltonian $H^{\text{SP}}$ is off-diagonal only in HF band index. This includes the continuum model Hamiltonian $H^{\text{CM}}$, the external sublattice potential $H^{\Delta}$, as well as a contribution $H^\text{scr}$ that accounts for the potential of the filled remote valence bands and subtraction of double-counted interactions relative to a fixed reference density (chosen to correspond to decoupled neutral graphene sheets~\cite{xie2020,kwan2020domain}).
	
	The interaction piece is
	\begin{align}
	A^\text{int}_{\bm{k}x;\bm{k'}y}(\bm{q})&=\delta_{\bm{k}\bm{k'}}\sum_{\bm{p}a}V^{\bm{p}K\downarrow a;\bm{k}K\downarrow y}_{\bm{p}K\downarrow a;\bm{k}K\downarrow x}-\delta_{\bm{kk'}}\delta_{xy}\sum_{\bm{p}}V^{\bm{k}+\bm{q},K'\downarrow+;\bm{p}K'\downarrow -}_{\bm{k}+\bm{q},K'\downarrow+;\bm{p}K'\downarrow -}\\
	&+\delta_{\bm{kk'}}\sum_{\bm{p}\tau\sigma a}\left(\delta_{xy}V^{\bm{k}+\bm{q},K'\downarrow+;\bm{p}\tau\sigma a}_{\bm{p}\tau\sigma a;\bm{k}+\bm{q},K'\downarrow +}-V^{\bm{k}K\downarrow y;\bm{p}\tau\sigma a}_{\bm{p}\tau\sigma a;\bm{k}K\downarrow x}\right)n_{\tau\sigma a}\\
	&-V^{\bm{k}+\bm{q},K'\downarrow +;\bm{k'}K\downarrow y}_{\bm{k}K\downarrow x;\bm{k'}+\bm{q},K'\downarrow +}
	\end{align}
	where $V^{\alpha;\beta}_{\gamma;\delta}=\bra{\alpha;\beta}\hat{V}\ket{\delta;\gamma}$ is the interaction matrix element and $n_{\tau\sigma a}$ is 1 if the band is filled and 0 otherwise. The first line above reflects the loss of exchange energy of the created hole and the gain of exchange energy of the added electron. The second line accounts for the interaction of the PH pair with the filled active bands. The third line describes the mutual interaction of the electron and the hole.
	
	The corresponding expressions for the spin-flip exciton are
	\begin{align}
	Q^\dagger_\nu(\bm{q})&=\sum_{\bm{k}a}X^\nu_{\bm{k}a}(\bm{q})d^\dagger_{\bm{k}+\bm{q},K'\downarrow+}d^{\phantom{\dagger}}_{\bm{k},K'\uparrow a}\\
	A^{\text{SP}}_{\bm{k}x;\bm{k'}y}(\bm{q})&=\delta_{\bm{k}\bm{k'}}(H^\text{SP}_{\bm{k}+\bm{q},K'\downarrow; ++}\delta_{xy}-H^\text{SP}_{\bm{k},K'\uparrow;yx})\\
	A^\text{int}_{\bm{k}x;\bm{k'}y}(\bm{q})&=\delta_{\bm{k}\bm{k'}}\sum_{\bm{p}a}V^{\bm{p}K'\uparrow a;\bm{k}K'\uparrow y}_{\bm{p}K'\uparrow a;\bm{k}K'\uparrow x}-\delta_{\bm{kk'}}\delta_{xy}\sum_{\bm{p}}V^{\bm{k}+\bm{q},K'\downarrow+;\bm{p}K'\downarrow -}_{\bm{k}+\bm{q},K'\downarrow+;\bm{p}K'\downarrow -}\\
	&+\delta_{\bm{kk'}}\sum_{\bm{p}\tau\sigma a}\left(\delta_{xy}V^{\bm{k}+\bm{q},K'\downarrow+;\bm{p}\tau\sigma a}_{\bm{p}\tau\sigma a;\bm{k}+\bm{q},K'\downarrow +}-V^{\bm{k}K'\uparrow y;\bm{p}\tau\sigma a}_{\bm{p}\tau\sigma a;\bm{k}K'\uparrow x}\right)n_{\tau\sigma a}\\
	&-V^{\bm{k}+\bm{q},K'\downarrow +;\bm{k'}K'\uparrow y}_{\bm{k}K'\uparrow x;\bm{k'}+\bm{q},K'\downarrow +}.
	\end{align}

	\section{Exciton Berry Curvature}	
	In this section, we review how to characterize the curvature~\cite{yao2008} and topology of exciton bands.
	For the sake of notation, it will be useful to recap how it works in standard SP bands. In the discussion below we consider just two SP bands for simplicity. The generalization to multiple filled bands is straightforward. We assume for simplicity that each UC has a finite-orbital basis (indexed by $\alpha,\beta$). SP Bloch states are defined
	\begin{equation}
	\ket{\phi_{\bm{k},\tau}}=\sum_\alpha u_{\bm{k},\tau\alpha}\ket{\bm{k},\alpha}=\frac{1}{\sqrt{N}}\sum_{\bm{R},\alpha}u_{\bm{k},\tau\alpha}e^{i\bm{kR}}\ket{\bm{R},\alpha}
	\end{equation}
	where $\tau$ is a band index. We can only compare states with the same boundary conditions~\cite{resta2000}, so we consider the cell-periodic $u$'s
	\begin{equation}
	\ket{u_{\bm{k},\tau}}\equiv e^{-i\bm{k\hat{r}}}\ket{\phi_{\bm{k},\tau}}=\frac{1}{\sqrt{N}}\sum_{\bm{R},\alpha}u_{\bm{k},\tau\alpha}\ket{\bm{R},\alpha}.
	\end{equation}
	The connection and curvature are then
	\begin{gather}
	\bm{a}_\tau(k)=-i\sum_\alpha u^*_{\bm{k},\tau\alpha}\partial_{\bm{k}}u_{\bm{k},\tau\alpha}\\
	f_{\tau}(\bm{k})=\partial_{k_x}a_{\tau,y}(\bm{k})-\partial_{k_y}a_{\tau,x}(\bm{k}).
	\end{gather}	
	This same curvature is probed numerically by evaluating gauge-invariant fluxes~\cite{fukui2005}
	\begin{equation} f_\tau(\bm{k})\sim\text{Im}\log\braket{u_{\bm{k},\tau}|u_{\bm{k}+\hat{2},\tau}}\braket{u_{\bm{k}+\hat{2},\tau}|u_{\bm{k}+\hat{1}+\hat{2},\tau}}\braket{u_{\bm{k}+\hat{1}+\hat{2},\tau}|u_{\bm{k}+\hat{1},\tau}}\braket{u_{\bm{k}+\hat{1},\tau}|u_{\bm{k},\tau}},
	\end{equation}
	the integral of which is the Chern number. In the case that there are bands that are not completely energetically isolated, the above can be generalized to compute the Chern index of the set of bands~\cite{fukui2005}.
	
	Consider the following exciton wavefunction $\ket{\text{exc},\bm{q}}=\sum_{\bm{k}}\tilde{\psi}_{\bm{q}}(\bm{k})d^\dagger_{\bm{k}+\bm{q},-}d_{\bm{k},+}\ket{G}$, where $\ket{G}$ consists of a filled $+$ band. There are now two important steps. First, we perform a particle-hole transformation in the $\tau=-$ band so that we can work with two-particle states. Second, we need to be careful about CM and relative decomposition. The point is that the CM momentum should be defined in order to properly couple to the CM position $\bm{R}=\frac{\bm{r}_1+\bm{r}_2}{2}$. Therefore we define the exciton state as 
	\begin{equation}
	\ket{\psi^\text{exc}_{\bm{q}}}=\sum_{\bm{k}}{\psi}_{\bm{q}}(\bm{k})\ket{\phi_{\bm{k}+\frac{\bm{q}}{2},-}}\ket{\phi^*_{\bm{k}-\frac{\bm{q}}{2},+}}
	\end{equation}
	where $\psi_{\bm{q}}(\bm{k})=\tilde{\psi}_{\bm{q}}(\bm{k}-\frac{\bm{q}}{2})$. Now for a finite-orbital basis
	\begin{gather}
	\ket{\psi^\text{exc}_{\bm{q}}}=\sum_{\bm{k}}\psi_{\bm{q}}(\bm{k})\sum_{\alpha\beta \bm{r}_1\bm{r}_2}u_{\bm{k}+\frac{\bm{q}}{2},-\alpha}u_{\bm{k}-\frac{\bm{q}}{2},+\beta}^*e^{i(\bm{k}+\frac{\bm{q}}{2})r_1}e^{-i(\bm{k}-\frac{\bm{q}}{2})\bm{r}_2}\ket{\bm{r}_1,\alpha}\ket{\bm{r}_2,\beta}\\
	\ket{u^\text{exc}_{\bm{q}}}\equiv e^{-i\bm{q}(\frac{\hat{\bm{r}_1}+\hat{\bm{r}_2}}{2})}\ket{\psi_{\bm{q}}^{\text{exc}}}=\sum_{\bm{k}}\psi_{\bm{q}}(\bm{k})\sum_{\alpha\beta \bm{r}_1\bm{r}_2}u_{\bm{k}+\frac{\bm{q}}{2},-\alpha}u_{\bm{k}-\frac{\bm{q}}{2},+\beta}^*e^{i\bm{k}(\bm{r}_1-\bm{r}_2)}\ket{\bm{r}_1,\alpha}\ket{\bm{r}_2,\beta},
	\end{gather}
	where the $u$'s are defined so that they have the same CM boundary conditions for different $\bm{q}$. Hence a sensible connection can be defined analogously to the single-particle case
	\begin{align}
	\bm{a}^\text{exc}(\bm{q})&\equiv -i\bra{u_{\bm{q}}^\text{exc}}\partial_{\bm{q}}\ket{u_{\bm{q}}^\text{exc}}\\
	&=-i\sum_{\bm{k}}\psi_{\bm{q}}^*(\bm{k})\partial_{\bm{q}}\psi_{\bm{q}}(\bm{k})+\frac{1}{2}\sum_{\bm{k}}|\psi_{\bm{q}}(\bm{k})|^2a_-(\bm{k}+\frac{\bm{q}}{2})+\frac{1}{2}\sum_{\bm{k}}|\psi_{\bm{q}}(\bm{k})|^2a_+(\bm{k}-\frac{\bm{q}}{2}).
	\end{align}
	The exciton curvature is
	\begin{align}
	f^\text{exc}(\bm{q})\equiv&\partial_{q_x}a_y^\text{exc}(\bm{q})-\partial_{q_y}a_x^\text{exc}(\bm{q})\\
	=&-i\sum_{\bm{k}}\bigg[\big(\partial_{q_x}\psi^*_{\bm{q}}(\bm{k})\big)\big(\partial_{q_y}\psi_{\bm{q}}(\bm{k})\big)-\big(\partial_{q_y}\psi^*_{\bm{q}}(\bm{k})\big)\big(\partial_{q_x}\psi_{\bm{q}}(\bm{k})\big)\bigg]\\
	&+\frac{1}{4}\sum_{\bm{k}}|\psi_{\bm{q}}(\bm{k})|^2f_-(\bm{k}+\frac{\bm{q}}{2})-\frac{1}{4}\sum_{\bm{k}}|\psi_{\bm{q}}(\bm{k})|^2f_+(\bm{k}-\frac{\bm{q}}{2})\\
	&+\frac{1}{2}\sum_{\bm{k}}\bigg[\big(\partial_{q_x}|\psi_{\bm{q}}(\bm{k})|^2\big)\big(a_{-,y}(\bm{k}+\frac{\bm{q}}{2})\big)+\big(\partial_{q_x}|\psi_{\bm{q}}(\bm{k})|^2\big)\big(a_{+,y}(\bm{k}-\frac{\bm{q}}{2})\big)-(x\leftrightarrow y)\bigg]
	\end{align}
	It can be verified that the curvature is gauge-invariant under the following transforms
	\begin{gather}
	\ket{\phi_{\bm{k},-}}\rightarrow e^{i\varphi_-(\bm{k})}\ket{\phi_{\bm{k},-}}\\
	\ket{\phi_{\bm{k},+}}\rightarrow e^{i\varphi_+(\bm{k})}\ket{\phi_{\bm{k},+}}\\
	\psi_{\bm{q}}(\bm{k})\rightarrow e^{i\big(\varphi_+(\bm{k}-\frac{\bm{q}}{2})-\varphi_-(\bm{k}+\frac{\bm{q}}{2})\big)}e^{i\theta(\bm{q})}\psi_{\bm{q}}(\bm{k}).
	\end{gather}
	Numerically the curvature is probed by evaluating gauge-invariant loops, similarly to the SP case
	\begin{equation} f(\bm{k})\sim\text{Im}\log\braket{u^\text{exc}_{\bm{k}}|u^\text{exc}_{\bm{k}+\hat{2}}}\braket{u^\text{exc}_{\bm{k}+\hat{2}}|u^\text{exc}_{\bm{k}+\hat{1}+\hat{2}}}\braket{u^\text{exc}_{\bm{k}+\hat{1}+\hat{2}}|u^\text{exc}_{\bm{k}+\hat{1}}}\braket{u^\text{exc}_{\bm{k}+\hat{1}}|u^\text{exc}_{\bm{k}}}.
	\end{equation}

\end{appendix}
	
\end{document}